\documentclass[traditabstract,a4paper]{aa}
\usepackage{amsmath}
\usepackage{graphicx}
\usepackage{color}
\usepackage{natbib}

\bibpunct{(}{)}{;}{a}{}{,}

\begin{document}

\author{Edo van Uitert \inst{\ref{inst1},\ref{inst2}} \and Henk Hoekstra \inst{\ref{inst1}} \and Marijn Franx \inst{\ref{inst1}} \and David G. Gilbank \inst{\ref{inst3}} \and Michael D. Gladders \inst{\ref{inst4}} \and H.K.C. Yee \inst{\ref{inst5}}}
\institute{Leiden Observatory, Leiden University, Niels Bohrweg 2, NL-2333 CA Leiden, The Netherlands, email: vuitert@strw.leidenuniv.nl \label{inst1}\and  Argelander-Institut f\"ur Astronomie, Auf dem H\"ugel 71, 53121 Bonn, Germany \label{inst2} \and South African Astronomical Observatory, PO Box 9, Observatory 7935, South Africa \label{inst3}\and Department of Astronomy and Astrophysics, University of Chicago, 5640 S. Ellis Ave., Chicago, IL 60637, USA \label{inst4} \and Department of Astronomy and Astrophysics, University of Toronto, 50 St. George Street, Toronto, Ontario, M5S 3H4, Canada \label{inst5} }

\title{Stellar mass versus velocity dispersion as tracer of the lensing signal around bulge-dominated galaxies}

\titlerunning{Stellar mass versus velocity dispersion}

\abstract {We present the results of a weak gravitational lensing analysis to determine whether the stellar mass or the velocity dispersion is more closely related to the amplitude of the lensing signal around galaxies - and hence to the projected distribution of dark matter. The lensing signal on scales smaller than the virial radius corresponds most closely to the lensing velocity dispersion in the case of a singular isothermal profile, but is on larger scales also sensitive to the clustering of the haloes. We select over 4000 lens galaxies at a redshift $z<0.2$ with concentrated (or bulge-dominated) surface brightness profiles from the $\sim$300 square degree overlap between the Red-sequence Cluster Survey 2 (RCS2) and the data release 7 (DR7) of the Sloan Digital Sky Survey (SDSS). We consider both the spectroscopic velocity dispersion and a model velocity dispersion (a combination of the stellar mass, the size and the S\'{e}rsic index of a galaxy). Comparing the model and spectroscopic velocity dispersion we find that they correlate well for galaxies with concentrated brightness profiles. We find that the stellar mass and the spectroscopic velocity dispersion trace the amplitude of the lensing signal on small scales equally well. The model velocity dispersion, however, does significantly worse. A possible explanation is that the halo properties that determine the small-scale lensing signal - mainly the total mass - also depend on the structural parameters of galaxies, such as the effective radius and S\'{e}rsic index, but we lack data for a definitive conclusion.}


\maketitle

\section{Introduction}  
\hspace{4mm} Galaxies form and evolve in the gravitational potentials of large dark matter haloes. The physical processes that drive galaxy formation cause correlations between the properties of the galaxies and their dark matter haloes. Hence to gain insight into these processes, various properties of galaxies  (e.g. colour, metallicity, stellar mass, luminosity, velocity dispersion) can be observed and compared \citep[e.g.][]{Smith09,Graves09}. This has lead to the discovery of a large number of empirical scaling laws, such as the Faber-Jackson relation \citep{Faber76}. These scaling laws help us to disentangle the processes that govern galaxy formation, and serve as important constraints for the theoretical and numerical efforts in this field. Although much progress has been made over the last few decades, many details are still unclear and warrant further investigation. \\
\indent One key parameter in galaxy formation is thought to be the total mass of a galaxy. Galaxies that have more massive dark matter haloes than others attract more baryons as well, consequently form more stars which results in larger stellar masses.  The relation between the stellar mass and the total mass of galaxies has been studied with observations \citep[e.g.][]{Mandelbaum06,VanUitert11,Leauthaud11,More11,Wake11}, abundance-matching techniques \citep[e.g.][]{Behroozi10,Guo10,Moster10}, semi-analytical modeling \citep[e.g.][]{Somerville08,Zehavi12} and hydrodynamical simulations \citep[e.g.][]{Keres09,Crain09,Gabor11,Munshi12}, and the two components are indeed found to be correlated. Another property of galaxies that is related to the total mass is the velocity dispersion, the luminosity weighted dispersion of the motions of stars along the line-of-sight within a spectroscopic aperture. The velocity dispersion provides a dynamical estimate of the central mass, and correlates with the stellar mass \citep{Taylor10} and the total mass of galaxies \citep{VanUitert11}. \\
\indent A fundamental question that is of interest in this context is which property of galaxies is most tightly correlated to the total mass. This is interesting, because it shows which property in the centre of dark matter haloes is most intimately linked to the large-scale potential, and is therefore least sensitive to galaxy formation processes such as galaxy mergers and supernova activity that introduce scatter in these relations. The properties of galaxies we compare in this work are the stellar mass and the velocity dispersion. Note that there are various other observables that trace the total mass, and could have been used instead, but most of them are either expected to exhibit a large amount of scatter (e.g. metallicity), or they are closely related to the stellar mass (e.g. luminosity). \\
\indent The total mass of galaxies is not directly observable, and can only be determined by indirect means. An excellent tool to do this is via weak gravitational lensing. In weak lensing the distortion of the images of faint background galaxies (sources) due to the gravitational potentials of intervening structures (lenses) is measured. From this distortion, the differential surface mass density of the lenses can be deduced, which can be modelled to obtain the total mass. A major advantage of weak lensing is that it does not rely on physically associated tracers of the gravitational potential, making it a particular useful probe to study dark matter haloes of galaxies which can extend up to hundreds of kpcs, where such tracers are sparse. The major disadvantage of weak lensing is that the lensing signal of individual galaxies is too weak to detect as the induced distortions are typically 10-100 times smaller than the intrinsic ellipticities of galaxies. Therefore, the signal has to be averaged over hundreds or thousands of lenses to decrease the shape noise and yield a statistically significant signal. 
However, the average total mass for a certain selection of galaxies is still a very useful measurement, which can be compared to simulations. \\
\indent It is important to note that the lensing signal on small and large scales measures different properties of dark matter haloes. On projected separations larger than a few times the virial radius, the lensing signal is mainly determined by neighbouring structures, and therefore depends on the clustering properties of the lenses. Within the virial radius, on the other hand, the lensing signal traces the dark matter distribution of the halo that hosts the galaxy and is therefore directly related to the halo mass. In this work, we ignore the lensing signal at large scales and instead focus at the signal at small scales. \\
\indent This work is a weak-lensing analogy of the analysis presented in \citet{Wake12}, who performed a similar study using galaxy clustering instead of gravitational lensing. One of their main findings is that the spectroscopic velocity dispersion is more tightly correlated to the clustering signal than either the stellar mass or the dynamical mass. This implies that the velocity dispersion better traces the properties of the halo that determine its clustering, that is the halo mass or the halo age. As the small scale weak lensing signal measures the halo mass, it allows us to disentangle the possible explanations of the clustering results.  \\
\indent The outline of this work is as follows. In Section \ref{sec_analysis4}, we discuss the various steps of the lensing analysis: we start with a description of the lens selection, then provide a brief outline of the creation of the shape measurement catalogues, and finally discuss the lensing analysis. The measurements are shown in Section \ref{sec_res}, and we conclude in Section \ref{sec_conc}. Throughout the paper we assume a WMAP7 cosmology \citep{Komatsu11} with $\sigma_8=0.8$, $\Omega_{\Lambda}=0.73$, $\Omega_M=0.27$, $\Omega_b=0.046$ and $h_{70}=H_0/70$ km s$^{-1}$ Mpc$^{-1}$ with $H_0$ the Hubble constant. All distances quoted are in physical (rather than comoving) units unless explicitly stated otherwise.

\section{Lensing analysis \label{sec_analysis4}}
\hspace{4mm}  In this study we use the $\sim$300 square degrees of overlapping area between the Sloan Digital Sky Survey \citep[SDSS;][]{York00} and the Red-sequence Cluster Survey 2 \citep[RCS2;][]{Gilbank10}. We use the SDSS to obtain the properties of the lenses (e.g. stellar mass, velocity dispersion), information that is not available in the RCS2. The lensing analysis is performed on the RCS2, because it is $\sim$2 magnitudes deeper than the SDSS in $r'$. The increase in depth combined with a median seeing of 0.7$''$, which is a factor of two smaller than the seeing in the SDSS, results in a source galaxy number density that is about five times higher, and a source redshift distribution that peaks at $z$$\sim$0.7. Therefore, the RCS2 enables a high-quality detection of the lensing signal, even for a moderate number of lens galaxies.

\subsection{Lenses \label{sec_lenses}}
\hspace{4mm} The SDSS has imaged roughly a quarter of the entire sky, and has measured the spectra for about one million galaxies \citep{Eisenstein01,Strauss02}. The combination of spectroscopic coverage and photometry in five optical bands ($u,g,r,i,z$) in the SDSS provides a wealth of galaxy information that is not available from the RCS2. To use this information, but also benefit from the improved lensing quality of the RCS2, we use the 300 square degrees overlap between the surveys for our analysis. We match the RCS2 catalogues to the DR7 \citep{Abazajian09} spectroscopic catalogue, to the MPA-JHU DR7\footnotemark  
\footnotetext[1]{http://www.mpa-garching.mpg.de/SDSS/DR7/}
stellar mass catalogue and to the NYU Value Added Galaxy Catalogue (NYU-VAGC)\footnotemark
\footnotetext[2]{http://sdss.physics.nyu.edu/vagc/}
\citep{Blanton05,Adelman08,Padmanabhan08} which yields the spectroscopic redshifts, velocity dispersions, and the stellar masses of $1.7\times10^4$ galaxies. From these galaxies we select our lenses using criteria that are detailed below. \\
\begin{figure*}[t!]
  \resizebox{\hsize}{!}{\includegraphics[angle=-90]{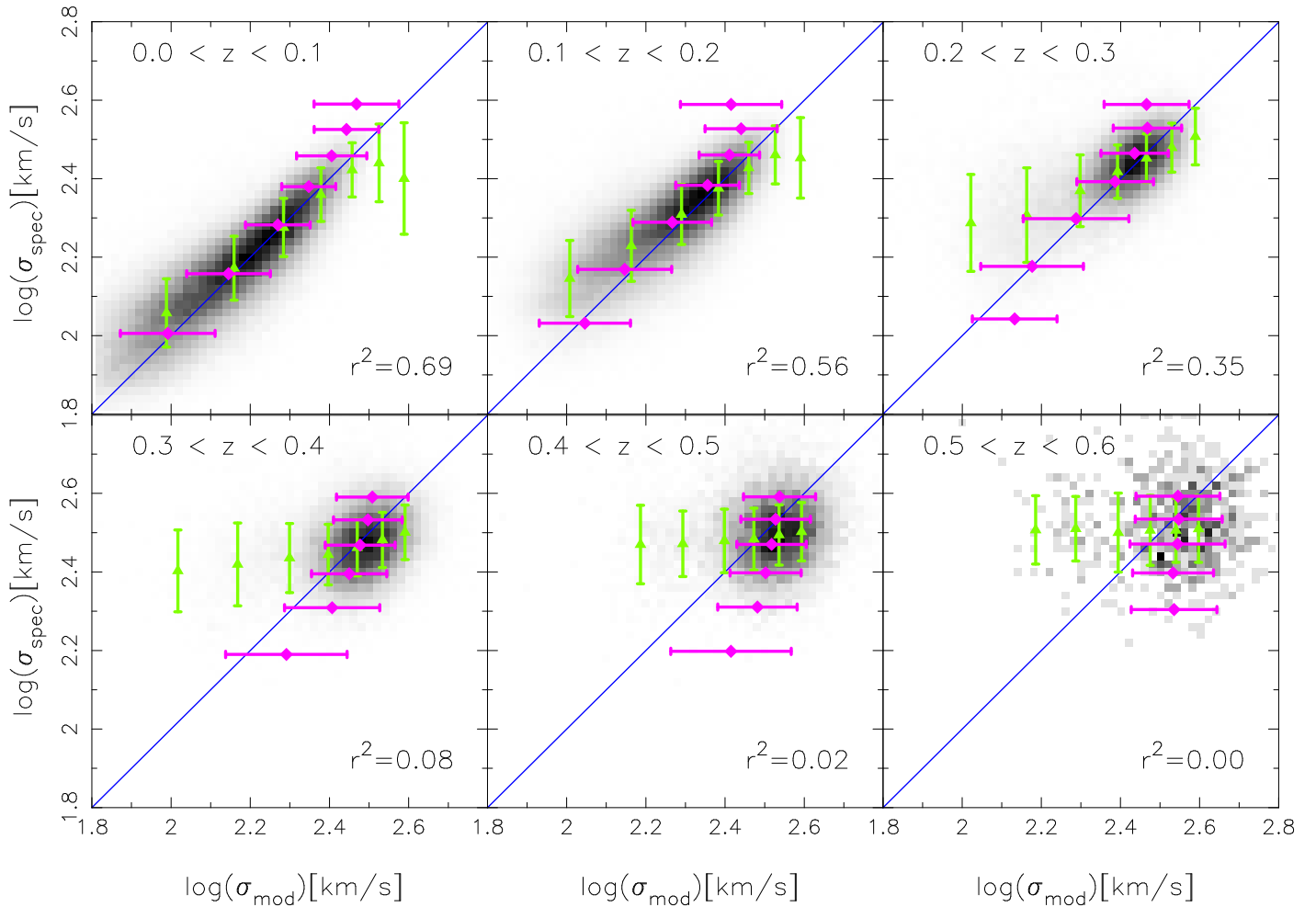}}
  \caption{Comparison of the spectroscopic velocity dispersions to the model velocity dispersions for all galaxies with SDSS spectroscopy. The green triangles show the average spectroscopic velocity dispersion for bins of model velocity dispersion, the purple diamonds show the average model velocity dispersion for bins of spectroscopic velocity dispersion. The error bars indicate the scatter. The blue line shows the one-to-one correspondence. Only galaxies with a spectroscopic velocity dispersion error smaller than 15\% have been used in the comparison. The velocity dispersions correlate well at $z<0.2$, but at $z>0.2$ the range in velocity dispersion becomes too small to assess whether this is still the case. The square of the correlation coefficient ${\rm{r}}^2$ of the galaxies in the range 1.8 $<\log_{10}(\sigma_{\rm{mod/spec}})<$ 2.8 km s$^{-1}$ is shown in the lower right corner of each panel. }
  \label{plot_sigsig}
\end{figure*}
\indent The spectroscopic fibre within which the velocity dispersion is measured has a fixed size. The physical region where the velocity dispersion is averaged is therefore different for a sample of galaxies with different sizes and redshifts. To account for this, we follow \citet{Bezanson11} and scale the observed spectroscopic velocity dispersion to a fixed size of $R_e/8$ using $\sigma_{\rm{spec}}=\sigma^{\rm{ap}}_{\rm{spec}}(8.0 r_{\rm{ap}}/R_e)^{0.066}$, with $r_{\rm{ap}}$=1.5$''$ the radius of the SDSS spectroscopic fiber, $R_e$ the effective radius in the $r$-band, and $\sigma^{\rm{ap}}_{\rm{spec}}$ the observed velocity dispersion. This correction is based on the best-fit relation determined using 40 galaxies in the SAURON sample \citep{Cappellari06}. However, the spectroscopic velocity dispersions provided in the DR7 spectroscopic catalogues are generally noisy for late-type galaxies. To obtain more robust velocity dispersion estimates for these galaxies, we also predict the velocity dispersion based on quantities that are better determined following \citet{Bezanson11}:
\begin{equation}
\sigma_{\rm{mod}}=\sqrt{\frac{GM_{*}}{0.557 K_V(n) R_e}},
\label{eq_sigmamod}
\end{equation}
with $M_*$ the stellar mass, $n$ the S\'{e}rsic index and $K_V(n)$ a term that includes the effects of structure on stellar dynamics, and can be approximated by \citep{Bertin02}
\begin{equation}
  K_V(n) \cong \frac{73.32}{10.465+(n-0.94)^2}+0.954.
\end{equation}
The equation for $\sigma_{\rm{mod}}$ is based on the results of \citet{Taylor10}, who demonstrated that the structure-corrected dynamical mass is linearly related to the stellar mass for a selection of low-redshift galaxies in the SDSS. \\
\indent The stellar mass estimates in the MPA-JHU DR7 catalogues are based on the model magnitudes. The S\'{e}rsic index and the effective radius in Equation (\ref{eq_sigmamod}), however, correspond to a different flux, i.e. the S\'{e}rsic model flux, which is the total flux of the best fit S\'{e}rsic model. This flux is also provided in the NYU-VAGC catalogue, and differs slightly from the model flux. To calculate $\sigma_{\rm{mod}}$ consistently, we therefore scale the stellar mass with the ratio of the model flux to the S\'{e}rsic model flux. \\
\indent \citet{Bezanson11} find that the model and the observed velocity dispersion correlate very well in the range $60$ km s$^{-1}<\sigma<300$ km s$^{-1}$, for galaxies in the redshift range $0.05<z<0.07$, and for a few galaxies with redshifts $1<z<2.5$. The SDSS spectroscopic sample extends to $z\sim0.5$, and therefore contains many more massive galaxies. To determine whether the velocity dispersions correlate well in this range too, we compare the dispersions for the complete SDSS spectroscopic sample in Figure \ref{plot_sigsig}. We find that the velocity dispersions agree well, though at $z>0.2$ the range in velocity dispersion becomes too small to assess whether the velocity dispersions are still correlated. This is reflected by the correlation coefficient of the log of the velocity dispersions of galaxies in the range 1.8 $<\log_{10}(\sigma_{\rm{mod/spec}})<$ 2.8 km s$^{-1}$ which we show in the corresponding panels. \\
\indent To study whether the spectroscopic velocity dispersion and the model velocity dispersion agree equally well for different galaxy types, we split the galaxies based on their {\it frac\_dev} parameter from the SDSS photometric catalogues. This parameter is determined by simultaneously fitting $frac\_deV$ times the best-fitting De Vaucouleur profile plus (1-$frac\_deV$) times the best-fitting exponential profile to an object's brightness profile. The {\it frac\_dev} parameter is therefore a measure of the slope (or concentration) of the brightness profile of a galaxy. In the following we will refer to lenses with {\it frac\_dev} $>$ 0.5 as galaxies with a surface brightness profile with a high concentration; these are bulge-dominated, which is typical for early-type galaxies. The lenses with {\it frac\_dev} $<$ 0.5 are referred to as those with a low concentration brightness profile; they are disk-dominated as is generally the case for late-type galaxies. We select all galaxies with redshifts $z<0.2$, and show the comparison in Figure \ref{plot_sigsig_type}.
\begin{figure}[t!]
  \resizebox{\hsize}{!}{\includegraphics[angle=-90]{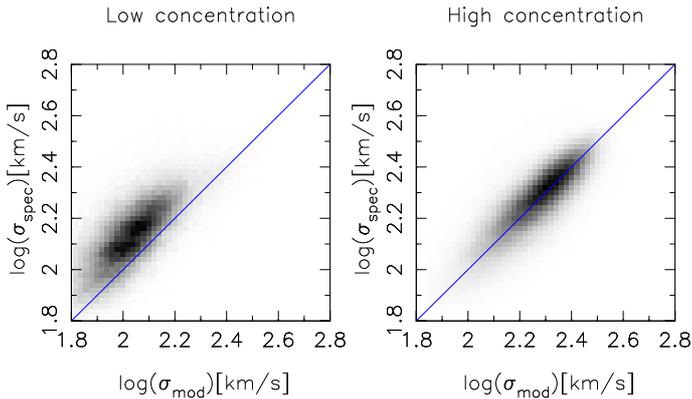}}
  \caption{Comparison of the spectroscopic velocity dispersions to the model velocity dispersions for galaxies with low concentration brightness profiles ({\it frac\_dev} $<$ 0.5) ({\it left}) and with high concentration brightness profiles ({\it frac\_dev} $>$ 0.5) ({\it right}) in the redshift range $0<z<0.2$. For the latter, the dispersions agree very well, but for the former, we find that the spectroscopic velocity dispersion is roughly 0.1 dex higher than the model velocity dispersion. }
  \label{plot_sigsig_type}
\end{figure}
We find that for the galaxies with high concentration (bulge-dominated) brightness profiles, the spectroscopic and model velocity dispersion agree very well. For those with low concentration brightness profiles, however, we find that the spectroscopic velocity dispersion is $\sim$0.1 dex higher than the model velocity dispersion. This is not surprising: \citet{Taylor10} found that the relation between the stellar mass and the structure-corrected dynamical mass has a weak dependence on the S\'{e}rsic index, i.e. the ratio of the stellar mass and the dynamical mass increases with increasing S\'{e}rsic index (see Figure 14 in Taylor et al. 2010). The offset in the relation between spectroscopic and model velocity dispersion for galaxies with low concentration brightness profiles is a direct consequence. It might be caused by the contribution of the disk velocity of spiral galaxies to the spectroscopic velocity dispersion. One could in principle apply a S\'{e}rsic index dependent correction, but we choose to use only galaxies with high concentration brightness profiles, because there are very few lenses with low concentration brightness profiles in the velocity dispersion range we are interested in. As a test we repeated the analysis including all lenses, and found that it did not affect our conclusions. \\
\indent In Figure \ref{plot_mstelsig}, we plot the spectroscopic and model velocity dispersion as a function of stellar mass. We only select galaxies with redshifts $z<0.2$; at higher redshifts, the range in velocity dispersions is too small to establish whether the correlation works well. The three lens samples we use are indicated by the dashed lines. We select all galaxies with high concentration brightness profile with a stellar mass $10.8<\log(M_*)<11.5$ in units of $h_{70}^{-1}M_\odot$; all with a model velocity dispersion 180 km s$^{-1}<\sigma_{\rm{mod}}<300$ km s$^{-1}$; and all with a spectroscopic velocity dispersion 180 km s$^{-1}<\sigma_{\rm{spec}}<300$ km s$^{-1}$ and a relative error of $<0.15$ in $\sigma_{\rm{spec}}$. With these criteria we select 4735, 4218 and 4317 lenses respectively, and they form the lens samples of this study. 
\begin{figure}[t!]
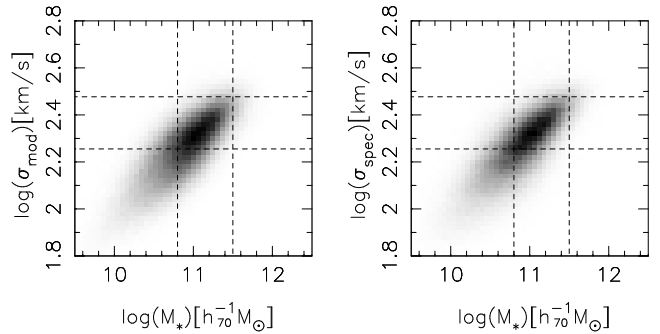

  \begin{minipage}[t]{0.48\linewidth}
      \includegraphics[width=1.\linewidth,angle=-90]{plot_mstelvssigmamod.ps}
  \end{minipage}
  \begin{minipage}[t]{0.48\linewidth}
      \includegraphics[width=1.\linewidth,angle=-90]{plot_mstelvssigmaspec.ps}
  \end{minipage}
  \caption{Model velocity dispersion ({\it left}) and  spectroscopic velocity dispersion ({\it right}) as a function of stellar mass. The dashed lines indicate the selection cuts for the lenses.}
  \label{plot_mstelsig}
\end{figure}

\subsection{Data reduction}
\hspace{4mm} The RCS2 is a nearly 900 square degree imaging survey in three bands ({\it g$'$, r$'$} and {\it z$'$}) carried out with the Canada-France-Hawaii Telescope (CFHT) using the 1 square degree camera MegaCam. The photometric calibration of the RCS2 is described in detail in \citet{Gilbank10}. The magnitudes are calibrated using the colours of the stellar locus and the overlapping Two-Micron All-Sky Survey (2MASS), and are accurate to $<0.03$ mag in each band compared to the SDSS. The creation of the galaxy shape catalogues is described in detail in \citet{VanUitert11}. We refer readers to that paper for more detail, and present here a short summary of the most important steps. \\
\footnotetext[3]{http://www.cfht.hawaii.edu/Instruments/Elixir/}
\footnotetext[4]{http://www1.cadc-ccda.hia-iha.nrc-cnrc.gc.ca/cadc/}
\indent We retrieve the Elixir\footnotemark \ processed images from the Canadian Astronomy Data Centre (CADC) archive\footnotemark. We use the THELI pipeline \citep{Erben05,Erben09} to subtract the image backgrounds, create weight maps that we use in the object detection phase, and to identify satellite and asteroid trails. To detect the objects in the images, we use {\tt SExtractor} \citep{BertinA96}. The stars that are used to model the PSF variation across the image are selected using size-magnitude diagrams. All objects larger than 1.2 times the local size of the PSF are identified as galaxies. We measure the shapes of the galaxies with the KSB method \citep{Kaiser95,LuppinoK97,Hoekstra98}, using the implementation described by \citet{Hoekstra98,Hoekstra00}. This implementation has been tested on simulated images as part of the Shear Testing Programme (STEP) 1 and 2 (the `HH' method in Heymans et al. 2006 and Massey et al. 2007 respectively), and these tests have shown that it reliably measures the unconvolved shapes of galaxies for a variety of PSFs. Finally, we correct the source ellipticities for camera shear, an instrumental shear signal which originates from slight non-linearities in the camera optics. The resulting shape catalogue of the RCS2 contains the ellipticities of 2.2$\times 10^7$ galaxies, from which we select the subset of approximately 1$\times 10^7$ galaxies that coincides with the SDSS. \\

\subsection{Lensing measurement \label{sec_shape}}
\hspace{4mm} In weak lensing studies, the ellipticities of the source galaxies are used to measure the azimuthally averaged tangential shear around the lenses as a function of projected separation:
\begin{equation}
  \langle\gamma_t\rangle(r) = \frac{\Delta\Sigma(r)}{\Sigma_{\mathrm{crit}}},
\end{equation}
where $\Delta\Sigma(r)=\bar{\Sigma}(<r)-\bar{\Sigma}(r)$ is the difference between the mean projected surface density enclosed by $r$ and the mean projected surface density at a radius $r$, and $\Sigma_{\mathrm{crit}}$ is the critical surface density:
\begin{equation}
  \Sigma_{\mathrm{crit}}=\frac{c^2}{4\pi G}\frac{D_s}{D_lD_{ls}},
\end{equation}
with $D_l$, $D_s$ and $D_{ls}$ the angular diameter distance to the lens, the source, and between the lens and the source, respectively. Since we lack redshifts for the background galaxies, we select galaxies with $22<m_{r'}<24$ that have a reliable shape estimate (ellipticities smaller than one, no SExtractor flag raised) as sources. We obtain the approximate source redshift distribution by applying identical magnitude cuts to the photometric redshift catalogues of the Canada-France-Hawaii-Telescope Legacy Survey (CFHTLS) ``Deep Survey" fields \citep{Ilbert06}. \\
\indent To correct the signal for systematic contributions, we compute the shear signal around a large number of random points to which identical image masks have been applied, and subtract that from the measured source ellipticities. Details on the calculation of this correction can be found in \citet{VanUitert11}. This correction effectively removes both the impact of residual systematics in the shape measurement catalogues, and the impact of image masks on tangential shear measurements. Note that this correction mostly affects large scales ($>$20 arcmin), as on small scales the lensing signal is generally averaged over many lens-source orientations causing the systematic contributions to average out. The source galaxy overdensity near the lenses is found to be a few percent at most, confirming that the lenses and sources barely overlap in redshift. Therefore, we do not have to correct the lensing signal for the contamination of physically associated galaxies in the source sample. \\
\indent Although neither the dark matter nor the baryonic component are well described by a singular isothermal sphere (SIS), the sum of the two components is remarkably close for massive elliptical galaxies on scales smaller than the effective radius \citep[e.g.][]{Treu04,Koopmans09}. 
The SIS signal is given by
\begin{equation}
  \gamma_{t,\mathrm{SIS}}(r) = \frac{r_E}{2r}=\frac{4\pi \sigma_{\rm{lens}}^2}{c^2}\frac{D_l D_{ls}}{D_s}\frac{1}{2r},
\end{equation}
where $r_E$ is the Einstein radius and $\sigma_{\rm{lens}}$ the lensing velocity dispersion. Van Uitert et al. (2011) show that for galaxies with 19.5 $<m_{r'}<$ 21.5 the SIS profile gives an accurate description of the lensing signal up to $\sim$300-500 $h_{70}^{-1}$ kpc. Based on the range of stellar masses and velocity dispersions of our lenses, we expect the majority of lenses to be central galaxies (see, e.g. van Uitert et al. 2011  or Mandelbaum et al. 2006 for estimates of the satellite fraction for galaxies in these ranges) that are considerably more massive than the average galaxy. Hence the range over which the lensing signal is well described by an SIS is likely even larger for our lenses. \\ 

\indent To determine whether the stellar mass or the velocity dispersion is a better tracer of the amplitude of the lensing signal, we would ideally select lenses in a very narrow range in stellar mass, split those in a high and low velocity dispersion bin, and compare their lensing signals. A difference between the lensing signal of the low and high velocity dispersion would indicate a residual dependence on velocity dispersion. Similarly, we would like to select lenses in a very narrow range of velocity dispersion, split them in stellar mass and compare their signals. Comparing the lensing signals of these four bins would allow us to determine whether the stellar mass or the velocity dispersion is more closely related to the lensing signal on small scales - and hence to the projected distribution of dark matter. \\
\indent Unfortunately, we do not have a sufficient number of lenses for this approach. Instead, we have to select lenses that cover a larger range in stellar mass (and velocity dispersion). We cannot simply split the lenses in velocity dispersion and compare their lensing signals, because the stellar mass and velocity dispersion are correlated, and the high velocity dispersion bin also has a larger mean stellar mass. To account for this, we determine how the lensing signal scales with stellar mass, and remove this trend from the high and low velocity dispersion bins. We also determine how the lensing signal scales with velocity dispersion, and remove this trend from the high and low stellar mass bin. If the lensing signal of galaxies strongly depends on the velocity dispersion, but only weakly on stellar mass, we expect a clear positive difference between the high and low velocity dispersion bins after we removed the trend with stellar mass. At the same time, we should see only a very small difference between the high and low stellar mass bin after removing the trend with velocity dispersion. Hence by studying the differences in the residual lensing signals, we can tell which observable is more closely related to the lensing signal of galaxies.

\section{Results \label{sec_res}}
\begin{figure}[t!]
  \resizebox{\hsize}{!}{\includegraphics{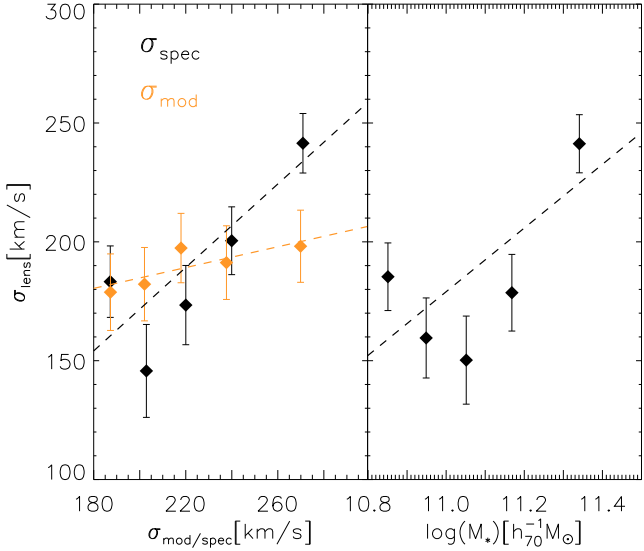}}
  \caption{Best-fit lensing velocity dispersion as a function of spectroscopic velocity dispersion ({\it left panel, black}), model velocity dispersion ({\it left panel, orange}) and stellar mass ({\it right panel}). Dashed lines indicate the best-fit linear relation between the observable and $\sigma_{\rm{lens}}$. The linear relations are used to remove the dependence of the lensing signal on these observables. }
  \label{plot_tracer_vsis}
\end{figure}
\begin{figure*}[t!]
  \resizebox{\hsize}{!}{\includegraphics{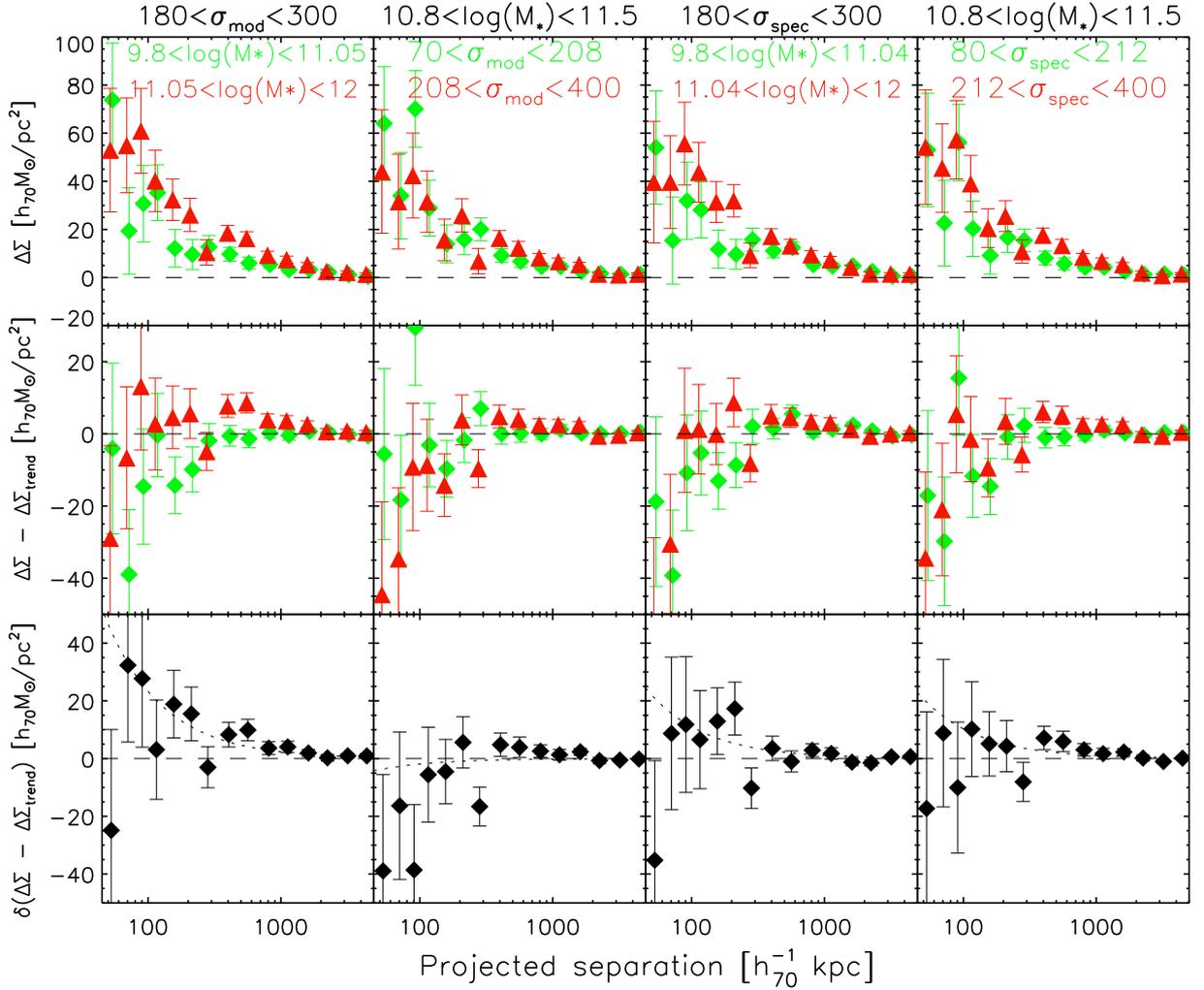}}
  \caption{In the top row, we show the lensing signal $\Delta \Sigma $ as a function of physical distance from the lens, for the lens samples that have been split by the median value of one of the observables, as indicated in the plots. Red triangles (green diamonds) indicate the signal of the lenses with larger (smaller) stellar masses/velocity dispersions. In the middle row, we show the lensing signal of the same samples after we subtracted the trend with the observable that is indicated on top of each column. The difference between the residual trends for the two lens samples are shown in the bottom row. A residual trend indicates that the lensing signal has a residual dependence on the observable indicated {\it inside} the corresponding panel of the first row, after removing the dependence on the observable indicated {\it on top} of that column. The dotted lines show the best-fit SIS profiles to the difference between the residuals. }
  \label{plot_mstelrmtrend}
\end{figure*}
\begin{figure}[t!]
  \resizebox{\hsize}{!}{\includegraphics{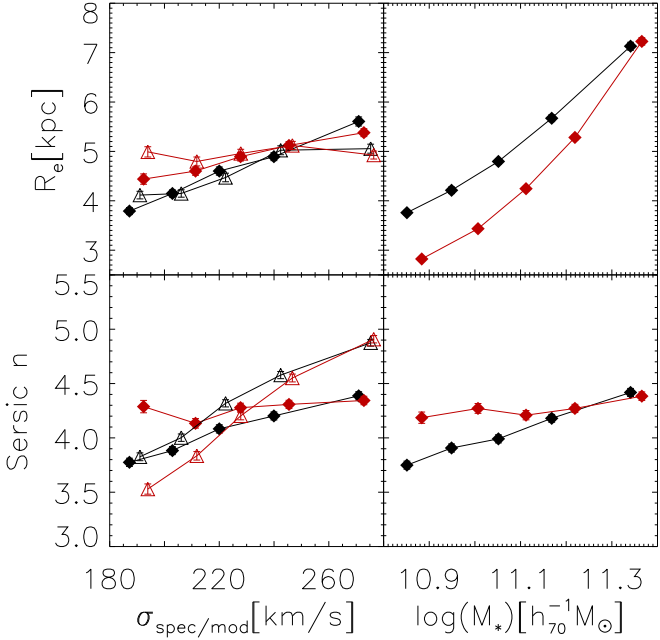}}
  \caption{Mean effective radius and S\'{e}rsic index as a function of spectroscopic velocity dispersion ({\it left, black filled diamonds}), model velocity dispersion ({\it left, black open triangles}) and stellar mass ({\it right, black diamonds}). In red, we show the averages for the lens samples that simultaneously satisfy 180 km s$^{-1}<\sigma_{\rm{mod}}<300$ km s$^{-1}$, 180 km s$^{-1}<\sigma_{\rm{spec}}<300$ km s$^{-1}$, $\delta\sigma_{\rm{spec}}/\sigma_{\rm{spec}}<0.15$ and 10.8 $h_{70}^{-1}M_\odot<\log(M_*)<$ 11.5 $h_{70}^{-1}M_\odot$. By simultaneously applying all selection criteria the average sizes and S\'{e}rsic indices of the samples change, which shows that we implicitly exclude galaxies from a certain area of structural parameter space.  }
  \label{plot_sigmstelnre}
\end{figure}
\begin{figure*}[t!]
  \resizebox{\hsize}{!}{\includegraphics[angle=-90]{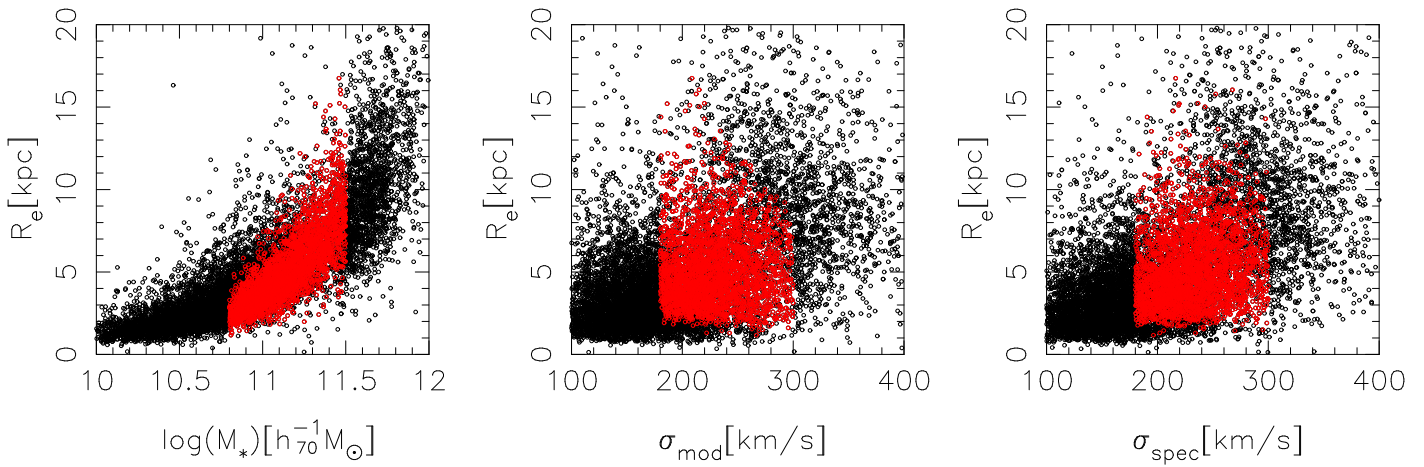}}
  \caption{The effective radius as a function of stellar mass ({\it left}), model velocity dispersion ({\it middle}) and spectroscopic velocity dispersion ({\it middle}). The black dots show all galaxies with high concentration brightness profiles with $z<0.2$, the red dots are the lenses that satisfy 180 km s$^{-1}<\sigma_{\rm{mod}}<300$ km s$^{-1}$, 180 km s$^{-1}<\sigma_{\rm{spec}}<300$ km s$^{-1}$, $\delta\sigma_{\rm{spec}}/\sigma_{\rm{spec}}<0.15$ and 10.8 $h_{70}^{-1}M_\odot<\log(M_*)<$ 11.5 $h_{70}^{-1}M_\odot$. By using all selection criteria simultaneously, we exclude large galaxies at low stellar masses and high model velocity dispersions, and small galaxies in the velocity dispersion samples. These selection biases could bias the lensing analysis if the lensing signal of a galaxy also depends on the effective radius.  }
  \label{plot_mstelvmre}
\end{figure*}

\hspace{4mm} To study whether the lensing signal mainly depends on stellar mass or velocity dispersion, we first have to determine how the lensing signal scales with these observables. We discuss how this is done for the model velocity dispersion; for the spectroscopic velocity and the stellar mass, we follow a similar approach. The general procedure is summarized below. \\
\begin{itemize}
\item[$\bullet$] We sort the lenses in model velocity dispersion, and divide them in five quintiles;
\item[$\bullet$] We measure the lensing signal of each quintile, to which we fit an SIS profile on scales between 50 $h_{70}^{-1}$kpc and 1 $h_{70}^{-1}$Mpc. This is roughly the range where the galaxy dark matter halo dominates the lensing signal. This results in five best-fit lensing velocity dispersions, $\sigma_{\rm{lens}}$;
\item[$\bullet$] We use the five values of $\sigma_{\rm{lens}}$ to fit the linear relation $\sigma_{\rm{lens}}=a_{\rm{mod}}\times(\sigma_{\rm{mod}}/200 \hspace{0.5mm}{\rm km \hspace{1mm} s}^{-1})+b_{\rm{mod}}$. We show the measurements and the fit in Figure \ref{plot_tracer_vsis}, and give the best-fit parameters in Table \ref{tab_fit};
\item[$\bullet$] We determine the median stellar mass of these lenses, and divide them into a low and high stellar mass sample. We measure the lensing signal of both samples, and show them in the top-left panel of Figure \ref{plot_mstelrmtrend};
\item[$\bullet$] For each lens in the low and high stellar mass sample, we use the model velocity dispersion to calculate $\sigma_{\rm{lens}}$ using the linear relation, and subtract their SIS profiles from the lensing signal. The residuals are shown in the middle-left panel of the same figure;
\item[$\bullet$] Finally, we determine the difference between the residual lensing signal of the high and low stellar mass bin, $\delta (\Delta \Sigma - \Delta \Sigma_{\rm{trend}})$, which is shown in the bottom-left panel.
\end{itemize}
\begin{table}
  \caption{Power law parameters}   
  \centering
  \begin{tabular}{c c c} 
  \hline
 &&\\
 $\sigma_{\rm{mod}}$ & $a_{\rm{mod}}$ & $b_{\rm{mod}}$ \\
 $[{\rm km/s}]$ & $[{\rm km/s}]$ & $[{\rm km/s}]$ \\
 &&\\
 $180<\sigma_{\rm{mod}}<300$ &  $44\pm48$ & $141\pm54$ \\
 $100<\sigma_{\rm{mod}}<400$ &  $93\pm25$ & $80\pm25$ \\
 && \\
 $\sigma_{\rm{spec}}$ & $a_{\rm{spec}}$ & $b_{\rm{spec}}$ \\
 $[{\rm km/s}]$ & $[{\rm km/s}]$ & $[{\rm km/s}]$ \\
 &&\\
 $180<\sigma_{\rm{spec}}<300$ &  $176\pm43$ & $-5\pm50$ \\
 $100<\sigma_{\rm{spec}}<400$ &  $129\pm21$ & $45\pm23$ \\
 &&\\
 $\log(M_*)$ & $a_{\rm{stel}}$ & $b_{\rm{stel}}$ \\
 $[h_{70}^{-1}M_\odot]$ & $[{\rm km/s}]$ & $[{\rm km/s}]$  \\
 &&\\
 $10.8<\log(M_*)<11.5$ &  $134\pm36$ & $179\pm8$ \\
 $10.5<\log(M_*)<12.0$ &  $118\pm22$ & $178\pm6$ \\
 &&\\
  \hline \\
  \end{tabular}
  \tablefoot{Best-fit power law slope $a_{\rm{x}}$ and offset $b_{\rm{x}}$ that describe the relation between galaxy property x (`mod', `spec' and `stel' for model velocity dispersion, spectroscopic velocity dispersion and stellar mass, respectively) and the lensing velocity dispersion for lens galaxies in the range that is indicated in the first column. Details of the fitting are described in the text. }
  \label{tab_fit}
\end{table}  
\hspace{4mm} When we subtract two SIS profiles with different amplitudes from each other, the result is also an SIS profile. Therefore, to quantify the residuals, we fit an SIS to $\delta (\Delta \Sigma - \Delta \Sigma_{\rm{trend}})$ on the same scales, and determine the residual Einstein radius, $r_{E}^{\rm{res}}$. These values can be found in Table \ref{tab_res}.\\
\indent Similarly, we determine the dependence of the lensing signal on spectroscopic velocity dispersion and stellar mass. For the spectroscopic velocity dispersion, we fit $\sigma_{\rm{lens}}=a_{\rm{spec}}\times(\sigma_{\rm{spec}}/200 \hspace{0.5mm}{\rm km \hspace{1mm} s}^{-1})+b_{\rm{spec}}$ and for the stellar mass, we fit  $\sigma_{\rm{lens}}=a_{\rm{stel}}\times\log(M_*/10^{11}\hspace{0.5mm} h_{70}^{-1}M_\odot)+b_{\rm{stel}}$. The best-fit parameters are shown in Table \ref{tab_fit}. These trends are removed from the lensing signals, and the residuals are shown in Figure \ref{plot_mstelrmtrend} (middle panel).  \\
\begin{table}
  \caption{Residual Einstein radii}   
  \centering
  \begin{tabular}{c c c c} 
  \hline
 &&&\\
 removed & residual  & $r_{E}^{\rm{res}}$ & $r_{E}^{\rm{res}}$\\
  trend&  dependence & [$h_{70}^{-1}$kpc] &[$h_{70}^{-1}$kpc]\\
 &&&\\
 $\sigma_{\rm{mod}}$ & $M_*$ & $0.88\pm0.25$ & ($0.78\pm0.25$) \\
 $M_*$ & $\sigma_{\rm{mod}}$ & $-0.18\pm0.24$ & ($-0.12\pm0.24$)  \\
 $\sigma_{\rm{spec}}$ & $M_*$ & $0.30\pm0.25$ & ($0.42\pm0.25$) \\
 $M_*$ & $\sigma_{\rm{spec}}$ & $0.37\pm0.24$ & ($0.42\pm0.24$) \\
 & & &\\
  \hline \\
  \end{tabular}
  \label{tab_res}
  \tablefoot{The residual Einstein radius, obtained by fitting an SIS profile to $\delta (\Delta \Sigma -\Delta \Sigma_{\rm{trend}}) $ between 50 $h_{70}^{-1}$kpc and 1 $h_{70}^{-1}$Mpc for a mean lens redshift of $z=0.13$. The bracketed values in the fourth column show the results for a different linear relation between the galaxy property and $\sigma_{\rm{lens}}$, as detailed in the text.}
\end{table}  
\indent In the bottom panel of the first column of Figure \ref{plot_mstelrmtrend}, we observe that after we have removed the lensing signal dependence on model velocity dispersion, $\delta (\Delta \Sigma - \Delta \Sigma_{\rm{trend}})$ is still positive on small scales, and therefore the lensing signal has a residual dependence on stellar mass. This residual dependence implies that the lensing signal still depends on stellar mass after removing its dependence on model velocity dispersion. In the panel next to it, where we have removed the dependence on stellar mass, we find that the difference between the residuals of the model velocity samples is consistent with zero, i.e. the lensing signal shows no residual dependence with model velocity dispersion. These trends are reflected by the values for $r_{E}^{\rm{res}}$ in Table \ref{tab_res}. The third and fourth columns of Figure \ref{plot_mstelrmtrend} show that if we remove the dependence on spectroscopic velocity dispersion, the difference of the residual signal of the high and low stellar mass sample is consistent with the difference between the residual signal of the high and low spectroscopic velocity dispersion samples after we removed the dependence on stellar mass. Note that even though the values of the residual Einstein radii change somewhat if we limit the analysis to smaller scales, which is likely mostly due to the larger statistical errors and because we probe different regions of the haloes, the trends we find do not change qualitatively. A quantitative characterization of the scale dependence of the signal could be performed with upcoming lensing surveys. \\
\indent Since we find that the lensing signal still depends on stellar mass after we remove its dependence on model velocity dispersion, but it does not depend on model velocity dispersion once we have removed the trend with stellar mass, our results suggest that the stellar mass is a better tracer of the lensing signal of galaxies than the model velocity dispersion. Furthermore, the stellar mass and the spectroscopic velocity dispersion trace the lensing signal equally well, as the residual Einstein radii are consistent. As a consistency check, we have also looked at the residual dependence on model velocity dispersion after removing the trend with spectroscopic velocity dispersion, and vice versa. These trends confirm our previous findings:  the spectroscopic velocity dispersion is more sensitive to the lensing signal of galaxies than the model velocity dispersion.\\
\indent There is a weak indication that the lensing signal has a residual dependence on stellar mass after we remove the trend with spectroscopic velocity dispersion, and vice versa. This would imply that both the stellar mass and the velocity dispersion contain independent information on the projected distribution of dark matter around galaxies. Unfortunately, we do not have sufficient signal-to-noise to obtain a clear detection.  \\
\indent The results depend on the linear relations we have fit to remove the dependence on the observables. To study how sensitive the residual trends are on these relations, we have also fit them using all galaxies with high concentration brightness profiles in the range 100 km s$^{-1}<\sigma_{\rm{spec}}<400$ km s$^{-1}$, 100 km s$^{-1}<\sigma_{\rm{mod}}<400$ km s$^{-1}$ and 10.5 $h_{70}^{-1}M_\odot<\log(M_*)<12$ $h_{70}^{-1}M_\odot$, respectively. The best-fit parameters of these fits are shown in Table \ref{tab_fit}. We repeated the analysis using these values, and show the residual Einstein radii between brackets in Table \ref{tab_res}. We find that this does not significantly change the results, i.e. the model velocity dispersion traces the lensing signal of galaxies worse than either the stellar mass or the spectroscopic velocity dispersion.\\
\indent It is somewhat surprising that $\sigma_{\rm{mod}}$ is a poorer tracer of the total mass than $\sigma_{\rm{spec}}$, particularly because we observe in Figure \ref{plot_sigsig} that they correlate well. A possible explanation is that they both trace the lensing velocity dispersion equally well on average, but with a different intrinsic scatter. To test this, we draw $10^6$ galaxies from the velocity dispersion function from \citet{Sheth03}, which we adopt as the lensing velocity dispersion $\sigma_{\rm{lens}}$. Next we assign them a spectroscopic and model velocity dispersion by drawing from a normal distribution with mean $\sigma_{\rm{lens}}$, using a larger scatter for $\sigma_{\rm{mod}}$ than for $\sigma_{\rm{spec}}$. Then we determine the average $\sigma_{\rm{lens}}$ in the five velocity dispersion bins, as in Figure \ref{plot_tracer_vsis}, and we compare the distribution of $\sigma_{\rm{mod}}$ and $\sigma_{\rm{spec}}$ as in Figure \ref{plot_sigsig_type}. Although we can choose the scatter of $\sigma_{\rm{mod}}$ and $\sigma_{\rm{spec}}$ such that we either reproduce Figure \ref{plot_sigsig_type} or Figure \ref{plot_tracer_vsis}, we cannot find a combination that reproduces both simultaneously. We also tested different distributions for $\sigma_{\rm{mod}}$ and $\sigma_{\rm{spec}}$ instead, and found similar results. Hence a difference in intrinsic scatter can at best only partly explain the different performance of $\sigma_{\rm{mod}}$ and $\sigma_{\rm{spec}}$ as tracers of the lensing signal. \\
\indent Another option is that there is another intrinsic property of galaxies with some additional dependence on the lensing signal. To study where the samples differ, we plot the average effective radii and S\'{e}rsic indices as a function $\sigma_{\rm{mod}}$ and $\sigma_{\rm{spec}}$ in Figure \ref{plot_sigmstelnre}. We find that at low velocity dispersions, the values of $r_e$ and $n$ are similar, but at high velocity dispersions, the lenses in the $\sigma_{\rm{spec}}$ samples have larger effective radii, whilst the lenses in the $\sigma_{\rm{mod}}$ have larger S\'{e}rsic indices. Hence the difference between the performance of $\sigma_{\rm{mod}}$ and $\sigma_{\rm{spec}}$ could be due to an additional dependence of the lensing signal on the structural parameters of the lenses. \\
\indent To test whether the lensing signal depends on the size of galaxies, we select the lenses from the model velocity dispersion sample, and remove the lensing signal dependence on $\sigma_{\rm{mod}}$. We then determine the median effective radius, split the lenses into a low and high effective radius sample and measure their residual lensing signal. As before, we measure the difference between the residual lensing signals of the high and low effective radius sample, to which we fit an SIS profile. We find $r_{E}^{\rm{res}}=0.53\pm0.25$ $h_{70}^{-1}$ kpc, which suggests that the lensing signal depends on the size of a galaxy. However, in Figure \ref{plot_sigmstelnre} and \ref{plot_mstelvmre} we find that the effective radius is correlated with stellar mass, so part of this residual may be caused by the dependence on stellar mass. Therefore, we repeat the test using the lenses from the stellar mass sample, and remove the lensing signal dependence on stellar mass. Then we determine the median effective radius, split the lenses into a low and high effective radius sample and measure the difference between their residual lensing signals. We find $r_{E}^{\rm{res}}=0.14\pm0.24$ $h_{70}^{-1}$ kpc. Studying the residual dependence on S\'{e}rsic index, we find $r_{E}^{\rm{res}}=0.04\pm0.25$ $h_{70}^{-1}$ kpc for the model velocity dispersion sample, and $r_{E}^{\rm{res}}=0.11\pm0.24$ $h_{70}^{-1}$ kpc for the stellar mass sample. These results do not provide conclusive evidence that the small-scale lensing signal depends on these structural parameters. \\
\indent Although the three lens samples overlap, they are not identical. Hence part of the trends we observe might actually be due to differences in the lens samples. To test this, we could define a fourth sample by selecting galaxies that pass all selection criteria, i.e. 180 km s$^{-1}<\sigma_{\rm{mod}}<300$ km s$^{-1}$, 180 km s$^{-1}<\sigma_{\rm{spec}}<300$ km s$^{-1}$, $\delta\sigma_{\rm{spec}}/\sigma_{\rm{spec}}<0.15$ and 10.8 $h_{70}^{-1} M_\odot<\log(M_*)<11.5$ $h_{70}^{-1} M_\odot$. However, if we simultaneously select on stellar mass and model velocity dispersion, we implicitly also select on effective radius and S\'{e}rsic index. This is demonstrated in Figure \ref{plot_sigmstelnre} and Figure \ref{plot_mstelvmre}, where we show the mean effective radius and S\'{e}rsic index for the lens samples. When we select lenses that pass all selection criteria, we exclude lenses with large effective radii and small S\'{e}rsic indices at low stellar mass, lenses with small effective radii and small S\'{e}rsic indices at low spectroscopic velocity dispersions, and lenses with small effective radii and large S\'{e}rsic indices at low model velocity dispersions. If the lensing signal of a galaxy also depends on its structural parameters, the lensing measurements of this fourth sample could be biased, making the results harder to interpret.\\
\indent \citet{Wake12} find that the ratio of the projected correlation functions of a high and low spectroscopic velocity dispersion sample at a fixed stellar mass is of the order 1.5-2 on scales $<$1$h^{-1}$ Mpc, whilst the ratio of a high and low stellar mass sample at a fixed velocity dispersion is close to unity. Our results show, however, that the lensing signal on small scales, which mainly depends on the halo mass, is traced equally well by the stellar mass and the spectroscopic velocity dispersion, and that potential differences are smaller than our statistical errors. The explanation could be that the difference in the amplitudes of the correlation functions are not mainly determined by how well they trace the halo mass. \citet{Wake12} mention two other possible causes: the relation between the velocity dispersion and halo age is tighter than between stellar mass and halo age, or tidal stripping of satellites which leads to a reduction in stellar mass, but does not affect the velocity dispersion. Although our results favour the latter two explanations, we cannot draw firm conclusions due to the low number of lens galaxies that could be used. \\
\indent It is important to note that we only study galaxies with high concentration brightness profiles, whilst the main results of \citet{Wake12} are based on a sample with mixed galaxy types. \citet{Wake12} do separate their sample based on colour and on morphology, and find some residual dependence on colour, but not on morphology, which suggests that their results would not have changed by much for a galaxy sample selected with similar criteria as in our work. We note, however, that the range of S\'{e}rsic indices of our galaxies is limited, as is shown in Figure \ref{plot_sigmstelnre}. Since S\'{e}rsic index and velocity dispersion generally correlate well, it might be that this is diluting the effect in our observations, which then would support the view that the spectroscopic velocity dispersion is a better tracer of the halo mass than the stellar mass. This could be investigated in more detail by repeating this analysis with a larger lens sample. 

\section{Conclusion \label{sec_conc}}
\hspace{4mm} In this work, we study which property of galaxies is most tightly correlated to the weak gravitational lensing signal on small scales for a sample of $\sim$4000 galaxies with high concentration brightness profiles ({\it frac\_dev} $>$ 0.5) at $z<0.2$. The properties we compare are the stellar mass, the spectroscopic velocity dispersion and the model velocity dispersion. We find that the lensing signal of galaxies is equally well traced by the stellar mass and the spectroscopic velocity dispersion. There is a weak indication for a residual dependence on stellar mass after removing the trend with spectroscopic velocity dispersion, and vice versa. This suggests that both tracers contain independent information on the projected distribution of dark matter around galaxies. Unfortunately, the signal-to-noise of our lensing measurements is not sufficient to make a definite statement.\\
\indent The model velocity dispersion traces the lensing signal significantly worse, which is surprising as the spectroscopic velocity dispersion and model velocity dispersion correlate well for our lenses. We find that this cannot be solely explained by assuming a larger intrinsic scatter of the model velocity dispersions compared to the spectroscopic ones. At high velocity dispersions, however, the lenses in the $\sigma_{\rm{mod}}$-sample have smaller effective radii and larger S\'{e}rsic indices than those in the $\sigma_{\rm{spec}}$-sample. This suggests that these structural parameters contain additional information on the projected distribution of dark matter around galaxies. To test this, we measure how the lensing signal depends on the size and S\'{e}rsic index of the lenses. We do not find conclusive evidence for a residual dependence on these structural parameters, which could be due to insufficient signal-to-noise caused by the relatively small lens sample of this study.\\
\indent The lensing signal on small projected separations from the lenses mainly depends on the halo mass. Our results therefore suggest that the stellar mass and spectroscopic velocity dispersion trace the halo mass equally well, but the model velocity dispersion does worse. However, at larger separations, neighbouring structures contribute to the lensing signal as well, and we cannot exclude the possibility that differences between the satellite fractions and large-scale clustering properties of the lens samples also have some effect.  \\
\indent Ideally, one should also remove the potential lensing signal dependence on the structural parameters of galaxies, i.e. split the lens sample both in velocity dispersion and structural parameters, and study the residual dependence on stellar mass. With the current data, we do not have sufficient signal-to-noise to perform such an analysis. However, we expect that at low-redshift ($z<0.2$) improvement is possible by repeating this analysis on the complete SDSS, while at higher redshifts the overlapping area between the RCS2 and the data release 9 \citep{SDSS9} of the SDSS could be used. Ultimately, one could simultaneously fit all these parameters, i.e. $M_h=f(\sigma,M_*,r_e,n,...)$, which could also contain products of the parameters such as $\sigma M_*$, and determine the covariance matrix between the coefficients. The relative magnitude of the coefficients would give new insights into which observables are important, and hence would provide valuable insights into galaxy formation processes. The expected lensing measurements from the space-mission Euclid \citep{Laureijs11} would be ideal for such a study. \\

\paragraph{Acknowledgements \\ \\} 
We would like to thank Pieter van Dokkum and David Wake for useful discussions, and Peter Schneider for a careful reading of the manuscript. HH and EvU acknowledge support from a Marie Curie International Reintegration Grant. HH is also supported by a VIDI grant from the Nederlandse Organisatie voor Wetenschappelijk Onderzoek (NWO). MDG thanks the Research Corporation for support via a Cottrell Scholars Award. The RCS2 project is supported in part by grants to HKCY from the Canada Research Chairs program and the Natural Science and Engineering Research Council of Canada. \\
\indent This work is based on observations obtained with MegaPrime/MegaCam, a joint project of CFHT and CEA/DAPNIA, at the Canada-France-Hawaii Telescope (CFHT) which is operated by the National Research Council (NRC) of Canada, the Institute National des Sciences de l'Univers of the Centre National de la Recherche Scientifique of France, and the University of Hawaii. We used the facilities of the Canadian Astronomy Data Centre operated by the NRC with the support of the Canadian Space Agency. 

\bibliographystyle{aa}

\end{document}